\documentclass[a4paper,11pt,fleqn]{article}
\usepackage{amsfonts,latexsym,epsfig}
\usepackage{color}

\newcommand{\R}{r}
\newcommand{\Rm}{r_{-1}}

\newcommand{\Gd}{\delta}

\newcommand{\Gl}{\lambda}
\newcommand{\GL}{\Lambda}

\def\moth{\mathsurround=0pt}
\newdimen\zo \zo=0pt

\def\tick{\leaders\hrule height 0.5ex depth 0pt \hskip 0.5pt}
\def\upboxfill{$\moth \setbox\zo\hbox{\tick}%
  \hskip 2pt\hbox to 0pt{$\tick$\hss}\hrulefill \hbox to 6pt{$\tick$\hss}$}

\def\dtick{\leaders\hrule height .34pt depth .5ex \hskip 0.5pt}
\def\downboxfill{$\moth \setbox\zo\hbox{\dtick}%
  \hskip 2pt\hbox to 0pt{$\dtick$\hss}\hrulefill \hbox to 6pt{$\dtick$\hss}$}

\newcommand{\cH}{{\cal H}}

\newcommand{\cM}{{\cal M}}

\newcommand{\BZ}{\mathbb{Z}}
\newcommand{\BR}{\mathbb{R}}

\newcommand{\be}{\begin{equation}}
\newcommand{\ee}{\end{equation}}
\newcommand{\ben}{\begin{displaymath}}
\newcommand{\een}{\end{displaymath}}
\newcommand{\ba}{\begin{eqnarray}}
\newcommand{\ea}{\end{eqnarray}}

\makeatletter
\@addtoreset{equation}{section}
\makeatother

\font\goth=eufm10

\begin{document}
\begin{titlepage}

\hfill{AEI-2001-025}

\hfill{IHES/P/01/12}

\hfill{LPT-ENS/01-14}

\hfill{ULB-TH/01-05}

\vspace{1cm}

\begin{centering}
                              
{\huge Hyperbolic Kac-Moody Algebras and Chaos in Kaluza-Klein Models}

\vspace{1.2cm}

{\Large Thibault Damour$^1$, Marc Henneaux$^{2,3}$, Bernard Julia$^{4}$
and Hermann Nicolai$^5$} \\

\vspace{.4cm}                                           
    
$^1$ Institut des Hautes Etudes Scientifiques,  35, route de
Chartres,  F-91440 Bures-sur-Yvette, France \\
\vspace{.2cm}

$^2$ Physique Th\'eorique et Math\'ematique,  Universit\'e Libre
de Bruxelles,  C.P. 231, B-1050, Bruxelles, Belgium  \\  
\vspace{.2cm}

$^3$ Centro de Estudios Cient\'{\i}ficos, Casilla 1469, Valdivia, Chile \\
\vspace{.2cm}

$^4$ Laboratoire de Physique Th\'eorique de l'Ecole Normale
Sup\'erieure, \\ 24, rue Lhomond, F-75231 Paris CEDEX 05 
\vspace{.2cm}

$^5$ Max-Planck-Institut f\"ur Gravitationsphysik, 
Albert-Einstein-Institut, 
M\"uhlenberg 1, D-14476 Golm, Germany \\

\vspace{1.5cm}
                      
\end{centering}

\begin{abstract}    

Some time ago, it was found that the never-ending oscillatory
chaotic  behaviour discovered by Belinsky, Khalatnikov and
Lifshitz (BKL) for the generic solution of the vacuum Einstein
equations in the vicinity of a spacelike (``cosmological")
singularity disappears in spacetime dimensions $D\equiv d+1>10$.  Recently,
a study of the generalization of the BKL chaotic behaviour to the
superstring effective Lagrangians has revealed that this chaos 
is rooted in the structure of the fundamental Weyl chamber of 
some underlying hyperbolic Kac-Moody algebra.  
In this letter, we show that the same connection applies to 
pure gravity in any spacetime dimension $\geq 4$,
where the relevant 
algebras are $AE_d$. In this way the disappearance of chaos in 
pure gravity models in $D \geq 11$ dimensions becomes linked to 
the fact that the Kac-Moody algebras $AE_d$ are no longer 
hyperbolic for $d \geq 10$. 

\end{abstract}

\vfill
\end{titlepage}
            
\section{Introduction}

A remarkable result in theoretical cosmology has been
the construction, by Belinsky, Khalatnikov and Lifshitz (BKL), 
of a generic solution to the 4-dimensional vacuum Einstein equations
in the vicinity of a spacelike (``cosmological") singularity \cite{BKL}.
This solution exhibits a never-ending oscillatory behaviour of the 
mixmaster type \cite{misner69,bkl69} with strong chaotic properties.
Some time ago, it was found that the BKL analysis for pure gravity 
leads to completely different qualitative features in 
spacetime dimensions $D \geq 11$ \cite{DHS,DHHST}. Namely,  
for those dimensions, the generic solution to the vacuum Einstein 
equations ceases to exhibit chaotic features, but is instead 
asymptotically characterized by a monotonic Kasner-like solution
(for a review, see \cite{DemDerHen}). The critical dimension $D=11$ 
was discovered by a straightforward but lengthy procedure,
with no direct interpretation. Another system for which chaos
is known to disappear is the pure gravity-dilaton system in
all spacetime dimensions \cite{BK,AR}.

More recently \cite{dh1,dh2,tDmH}, the BKL analysis was extended 
to the supergravity Lagrangians in $10$ \cite{sugra,GSW} and 11 dimensions
\cite{CJS} that emerge as the low energy limits of the superstring 
theories (IIA,IIB, I, HO, HE) and M Theory, respectively.
Contrary to what happens for the gravity-dilaton system in $10$ 
dimensions or pure gravity in $11$ dimensions, the chaotic oscillatory
behaviour was found to be generic in all superstring and $M$-theory 
models thanks to the $p$-forms present in the field spectrum \cite{dh1}.  
It was furthermore proved that this chaos was rooted in 
the structure of the fundamental Weyl chamber of some Kac-Moody 
algebra \cite{tDmH}.  More precisely, reformulating 
the asymptotic analysis of the dynamics as a billiard problem 
{\it \`a la} Chitre-Misner \cite{Chitre,Misner}, it was shown that 
the  never ending oscillatory BKL behaviour could be described 
as a relativistic billiard within a simplex in $9$-dimensional
hyperbolic space. The reflections on the faces of this billiard were
shown to generate a Coxeter group, which was then identified
with the Weyl group of the hyperbolic Kac-Moody algebras $E_{10}$ 
for the type IIA,IIB,and M theories, and $BE_{10}$ for the type 
I, HO, HE theories (for background on Kac-Moody algebras and notations, 
see the textbooks \cite{Kac,MP}). In this way, a relation was established
between the fact that the billiard has finite volume, and hence chaotic
dynamics, and the hyperbolicity of the underlying indefinite
Kac-Moody algebras $E_{10}$ and $BE_{10}$.  

In this letter, we re-examine the case of pure gravity in arbitrary 
spacetime dimension $D \equiv d+1$ in the light of these results. 
We demonstrate that the asymptotic dynamics (for
$t \rightarrow 0$, at any point in space)
can again be viewed as a 
billiard in the fundamental Weyl chamber of an 
indefinite Kac-Moody
algebra, which is now 
$AE_d  \equiv A_{d-2}^{\wedge \wedge} \equiv A_{d-2}^H$.
This algebra is the 
``overextended'' \cite{BJ} or ``canonical hyperbolic" extension \cite{Kac} of
the (finite dimensional) Lie algebra $A_{d-2}$; its associated
Dynkin diagram is obtained by attaching, at the affine node,
one more node to the 
Dynkin diagram of the affine algebra $A^{(1)}_{d-2} \equiv 
A_{d-2}^{\wedge}$ and is
displayed in Figure 1.
\begin{figure}
\begin{center}
\input{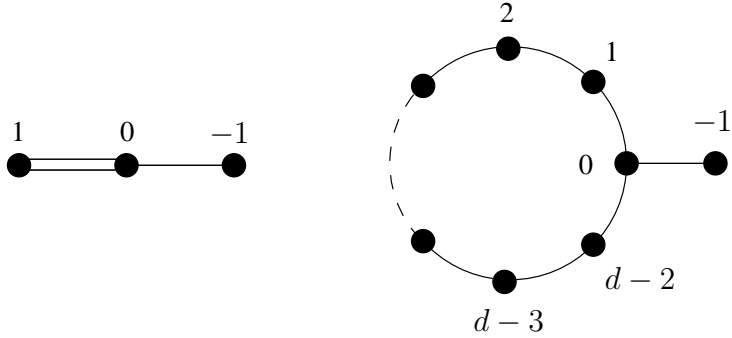}
\caption{Dynkin diagram of $AE_d$ (with $AE_3$ on the left)}
\end{center}
\end{figure}
The algebra $AE_3 \equiv A_{1}^{\wedge \wedge} \equiv A_1^H$
has been particularly studied in \cite{FF} and was related in \cite{HN}
to $D=4$ (super-)gravity.
Note that the double-line (in the conventions of \cite{Kac})
in its Dynkin diagram can be viewed as the formal limit
of the loop of $AE_d$ as $d \rightarrow 3$.
[It is interesting to remark that the Weyl group of $AE_3$ is $PGL_2(\BZ)$,
which is arithmetic \cite{Vinberg,FF}.]
Furthermore we show very explicitly how the occurrence of chaotic 
behaviour is correlated to the hyperbolicity of the underlying Kac-Moody
algebra. More specifically, the algebras $AE_d$ are hyperbolic (in the
sense defined in section 3 below) for $d<10$, 
whence pure gravity in dimensions $4\leq D \leq 10$ is chaotic, whereas 
chaos disappears in dimensions $D\geq 11$ in accord with the fact 
that the algebras $AE_d$ are no longer hyperbolic for $d \geq 10$.

The very existence of a connection between the BKL dynamics and 
indefinite Kac-Moody algebras is already remarkable in itself. For 
the generic Einstein system with matter couplings, one can always define
a billiard that describes the asymptotic dynamics, but in general,
this billiard will not exhibit any noticeable regularity properties.  
In particular, the faces of this billiard need not intersect at 
angles which are submultiples of $\pi$, and consequently the 
associated reflections will not generate a Coxeter (discrete)
reflection group
in general; {\it a fortiori}, the billiard need not be the fundamental 
Weyl chamber of any Kac-Moody algebra. The hyperbolic Kac Moody
algebra $E_{10}$ (and $DE_{10}$) was already conjectured in \cite{julia80,BJ}
to be a hidden symmetry of maximal supergravity reduced to one
dimension. The results of \cite{tDmH} and of this letter indeed support
the idea that hyperbolic Dynkin diagrams play a key r\^ole in the
massless bosonic sectors of supergravity and superstring theory. 
But we should emphasize that the 
Kac-Moody algebras do not appear in the present BKL analysis as symmetry
algebras with associated Noether charges.  They underlie nevertherless
the dynamics through their Weyl group,
in the sense that the dynamics
can be described in terms of ``Weyl words'' $W_{i_1} W_{i_2} \dots$
made out of the ``letters'' $W_i$ generating the Weyl reflections.

It is amazing to see the chaos
being controlled by the U-duality group $G$ of the toroidal compactification
to 3 dimensions via its overextension $G^{\wedge \wedge}$.  Recently,
it has been shown
\cite{HJK} that both $G=SO(8,8)$ and $G=SO(8,9)$ are the U-duality groups of
anomaly-free string models; in fact, other $SO(8,8+n)$ groups can be realised
beyond the heterotic $SO(8,24)$. A possible explanation for the
universality of $BE_{10}$ will be given there as well.

\section{Gravitational billiard in $d+1$ dimensions}

We first review how Einstein's theory gives rise to a ``gravitational 
billiard'' as one approaches a cosmological singularity; for more
details, see \cite{tDmH}. As usual, we assume that the singularity 
is at $t\rightarrow 0^+$, where $t$ is the proper time in a Gaussian 
coordinate system adapted to the singularity. In fact, 
it is convenient to use a time coordinate 
$\tau \sim - \log t$
such that $\tau \rightarrow + \infty$ as $t\rightarrow 0^+$ \cite{BKL,misner69}. 
In the asymptotic limit, the metric takes the form
\begin{equation}
ds^2 = - (N \sqrt{g} d\tau)^2 + \sum_{\mu =1}^d 
\exp {[- 2 \beta^\mu (\tau, x^i)]} \; 
(\omega^\mu)^2 \, , 
\label{ds2}
\end{equation}
where the time dependence of the spatial one-forms 
$\omega^\mu \equiv e^\mu_i(x^j, \tau) dx^i $ 
($i = 1 , \cdots, d$) 
can be neglected with respect
to the time-dependence of the scale functions $\beta^\mu$. 
In (\ref{ds2}), $N$ is the (rescaled) lapse $\sqrt{- g_{00}/g}$, where 
$g = \exp{(-2 \sum_{\mu=1}^d \beta^\mu)}$ is the determinant of
the spatial metric in the frame $\{ \omega^\mu \}$.
We assume $d \geq 3$ (i.e. $D\geq 4$) since pure gravity in $D = 3$
spacetime dimensions has no local degrees of freedom.  

The central feature that enables one to investigate  the equations 
of motion in the vicinity of a spacelike singularity is the asymptotic 
decoupling of the dynamics at the different spatial points \cite{BKL}. 
The remaining effect of the spatial gradients can be accounted for by
potential terms for the local scale factors $\beta^\mu$. Therefore,  
we focus from now on a specific spatial point and drop 
reference to the spatial coordinates $x^i$.
In the limit $\tau \rightarrow + \infty$, the dynamics
for the scale factors
$\beta^\mu$ is governed by the action
\begin{equation}
S[\beta^\mu(\tau), N(\tau)] = \int d\tau \left[
\frac{G_{\mu \nu}}{N} \frac{d \beta^\mu}{d \tau} \frac{d \beta^\nu}{d  \tau}
- N \, V(\beta^\mu) \right]
\label{action}
\end{equation}                    
where $G_{\mu \nu}$ is the metric defined by the Einstein-Hilbert
action in a $d$-dimensional auxiliary space $\cM_d$ spanned by
the ``coordinates'' $\beta^\mu$, which must not be confused 
with physical space-time. This metric is flat and of Minkowskian 
signature $(-, +, +, \cdots, +)$; explicitly, it reads
\begin{equation}\label{metric}
G_{\mu \nu} V^\mu W^\nu = \sum_{\mu = 1}^{d} V^\mu W^\mu
- \big(\sum_{\mu = 1}^{d}V^\mu \big) \big(\sum_{\nu = 1}^{d}W^\nu \big),
\end{equation}
We shall also need the inverse metric $G^{\mu \nu}$
\begin{equation}\label{inversemetric}
G^{\mu \nu} \theta_\mu \psi_\nu = \sum_{\mu = 1}^{d}\theta_\mu \psi_\mu
- \frac{1}{d-1} \big(\sum_{\mu = 1}^{d}\theta_\mu
\big) \big(\sum_{\nu = 1}^{d}\psi_\nu\big).
\end{equation}    
In (\ref{action}), the potential $V$ is a sum of sharp wall potentials,
\begin{equation}
V = \sum_i V_i,\; \; \; V_i = \Theta_\infty\big(- 2 w_i(\beta)\big)
\label{pot}
\end{equation}
where $\Theta_\infty$ vanishes for negative argument and 
is (positive) infinite for positive 
argument\footnote{Of course, the factor $2$ in the argument of 
$\Theta_\infty$ in (\ref{pot}) could be dropped
($\Theta_\infty( \lambda x) = \Theta_\infty(x)$ for $\lambda >0$),
but we keep it in order to emphasize that the walls come with a
natural normalization linked to the fact that they initially appear
as Toda walls $\sim \exp(-2 w_i(\beta))$ \cite{tDmH}.  These
exponential walls become sharp in
the Chitre-Misner limit \cite{Chitre,Misner}, generalized to higher
dimensions \cite{Ru,tDmH}.}. 
The functions $w_i(\beta)$ are homogeneous linear forms, viz. 
\begin{equation}
w_i(\beta)= w_{i \mu} \beta^\mu
\end{equation}
where the covectors $w_{i\mu}$ will be given explicitly below. 

Varying the rescaled lapse $N$ yields the Hamiltonian constraint
\begin{equation}
G_{\mu \nu} \frac{d \beta^\mu}{d \tau} \frac{d \beta^\nu}{d  \tau}
+ V = 0 \label{HC}
\end{equation}
where we have set $N=1$ (i.e., $dt = - \sqrt{g} d \tau$) after taking 
the variation, since this gauge choice simplifies the formulas 
(note that this implies indeed $\tau \sim - \log t$
since $\sqrt{g} \sim t$ \cite{BKL,misner69}). The dynamics is 
also subject to the spatial diffeomorphism (momentum) constraints,
but these affect the spatial gradients of the initial data and 
need not concern us here.    

We stress that the action (\ref{action}) is {\em not} obtained
by making a dimensional reduction to one dimension
of the $D$-dimensional Einstein-Hilbert
action assuming some internal $d$-dimensional group
manifold. 
Rather, 
the action (\ref{action}), or, more precisely, the sum over all
spatial points of copies of (\ref{action}), supplemented by the 
momentum constraints, is supposed to yield the asymptotic dynamics
in the limit $t \rightarrow 0^+$ for generic inhomogeneous
solutions \cite{BKL}. We should mention that the derivation 
of (\ref{action}) from the Einstein-Hilbert action involves a number 
of steps that have not been rigorously justified so far. 
Nevertheless, there is now a wealth of supporting evidence for the 
BKL analysis, both of analytical and of numerical type \cite{Berger,AlR}. 

Let us study the dynamics of the billiard ball whose motion
is described by the functions $\beta^\mu = \beta^\mu(\tau)$. From
(\ref{pot}) we immediately see that the interior region of the
billiard is defined by the inequalities $w_i(\beta)\geq 0$, and 
that its walls are coincident with the hyperplanes $w_i(\beta)=0$.
Away from the walls, the Hamiltonian constraint becomes
\begin{equation}
G_{\mu \nu} \frac{d \beta^\mu}{d \tau} \frac{d \beta^\nu}{d  \tau}
=0.
\end{equation}     
Thus the ball travels freely at the speed of light on straight lines 
until it hits one of the walls and gets reflected. The change
of the velocity $v^\mu \equiv \dot \beta^\mu$ after a collision
on the wall $w_i(\beta) = 0$ is given by a geometric reflection
in the corresponding wall hyperplane \cite{DHHST,tDmH}
\begin{equation}
v^\mu \rightarrow {v'}^\mu = \big(W_i (v)\big)^\mu
 \equiv v^\mu -2 \frac{w_{i\nu} v^\nu}{w_{i \rho}w^{\rho}_i}
w_i^\mu
\; \; \; \; (\hbox{no sum over $i$)}
\label{coll} 
\end{equation}
where $w_i^\mu \equiv G^{\mu \nu} w_{i \nu}$ are the contravariant
components of $w_i$.
For a timelike wall (whose normal vector is spacelike), the
reflection is an orthochronous Lorentz transformation;
hence the velocity remains null and future-oriented. 
Let ${\cal C}^+$ denote the future light cone with vertex at the origin 
($\beta^\mu = 0$) where
the walls intersect.  In the 
asymptotic regime under study, the initial point from which
one starts the motion has positive value of the timelike
combination $\sum_{\mu=1}^d \beta^\mu$ of the coordinates;
therefore, since the walls $w_i(\beta) = 0$ are all timelike
-- see below --, the ball wordline 
remains within ${\cal C}^+$ \cite{tDmH}.

The confinement of the billard motion to the forward light cone
enables one to project, if one so wishes, the piecewise
linear motion of the ball in the Minkowski space $\cM_d$ 
onto the upper sheet $\cH_{d-1}$ of the unit hyperboloid:
\begin{equation}
\cH_{d-1}: \; G_{\mu \nu} \beta^\mu \beta^\nu = -1, \; \; \;
\sum_{\mu = 1}^d  \beta^\mu >0.
\end{equation}
A projection is in fact physically necessary in order to
take into account the gauge redundancy (time-reparametrization 
invariance) and its associated Hamiltonian constraint. One of the 
$\beta^\mu$'s does not correspond to an independent degree 
of freedom.  The projection to the upper hyperboloid $\cH_{d-1}$ 
corresponds to viewing the $d-1$ coordinates of $\cH_{d-1}$
as the physical degrees of freedom and $\sum_{\mu = 1}^d  \beta^\mu$
(or a function of it) as the ``time" (see e.g. \cite{Ryan}).
For practical purposes, however, it is also convenient to keep the 
redundant description in terms of which the evolution is piecewise 
linear. We shall switch back and forth between the two descriptions. 
Note that the linear motion of $\beta^\mu$ projects to a geodesic 
motion on hyperbolic space $\cH_{d-1}$, so the problem is equivalent, 
in the limit under consideration, to a billiard in hyperbolic space.

We now wish to describe in more detail the convex (half) cone 
${\cal W}^+$ defined by the simultaneous fulfillment
of all the conditions $w_i(\beta) \geq 0$, 
to which the motion of the billiard ball is also confined. There are 
altogether two types of walls. Setting $n \equiv d-2$, they are
\begin{enumerate}
\item Symmetry walls \cite{tDmH}
\begin{eqnarray}
w_i (\beta) &=& \beta^{i} - \beta^{i-1} \;  \; \; \; 
(i = 2, \cdots, n \equiv d-2),
\label{symmetry1} \\
w_0 (\beta) &=& \beta^{d-1}- \beta^{d-2}, \label{symmetry2}
\\ w_{-1}(\beta) &=&
\beta^{d}- \beta^{d-1}
\label{symmetry3}
\end{eqnarray}
\item Gravitational wall \cite{DHS}
\begin{equation}
w_1(\beta) = 2 \beta^1 + \sum_{i=2}^{d-2} \beta^i \; \; \; (d \geq 4)
\label{gravitational}
\end{equation}
(for $d = 3$, $w_{-1} = \beta^3 - \beta^2$, $w_0 = \beta^2 -
\beta^1$ and $w_1 = 2 \beta^1 $).
\end{enumerate}                   
There is a total of $d$
walls, which are all timelike
since the associated wall forms (normal vectors) $w_i$ 
($i= -1, 0, 1, \cdots n$) are spacelike in any spacetime dimension:
\begin{equation}
G^{\mu \nu} w_{i \mu} w_{i \nu} = 2 \; \; (\hbox{$i$ fixed}).
\end{equation}  
The walls therefore intersect the upper light cone  ${\cal C}^+$.

The qualitative dynamics of the billiard can  be understood in
terms of the relative positions of ${\cal C}^+$ and ${\cal W}^+$.
Two cases are possible :
\begin{enumerate}
\item ${\cal W}^+$ is contained within ${\cal C}^+$ (i.e., all
vectors of ${\cal W}^+$ are timelike or null);
\item ${\cal W}^+$ is not entirely contained within ${\cal C}^+$
(i.e., there are not only timelike and null but also spacelike
vectors in ${\cal W}^+$).
\end{enumerate}
In the first case, the walls define a generalized, finite-volume simplex 
in hyperbolic space $\cH_{d-1}$ (generalized because some vertices
can be at infinity, which occurs when some edges of the cone ${\cal W}^+$
are lightlike\footnote{The edges of ${\cal W}^+$
are the (one-dimensional) intersections
of $d-1$ distinct faces of ${\cal W}^+$.}). 
As the walls are timelike, 
the ball will undergo an infinite number of collisions because, moving
at the speed of light, it will always catch up with one of the walls.  
The only exception, of measure zero, occurs when the ball moves 
precisely parallel to a lightlike edge of the billiard (there is
always at least one such edge). As we shall 
see in the next section, the dihedral angles of the wall are all 
submultiples of $\pi$, so that the reflections on the sides of the 
billiard generate a discrete group of isometries of hyperbolic space. 
Similarly to what happens in the superstring case \cite{tDmH}, the 
projected dynamics on $\cH_{d-1}$ is then chaotic (Anosov flow) 
according to general theorems on the geodesic motion on finite-volume 
manifolds with constant negative curvature.  

In the second case, some walls intersect outside ${\cal C}^+$ and 
the billiard on $\cH_{d-1}$ has infinite volume.
The ball undergoes a finite number of collisions until its motion
is directed toward a region of ${\cal W}^+$ that lies outside 
${\cal C}^+$.  It then never catches a wall anymore because it
cannot leave ${\cal C}^+$: no ``cushion" impedes
its motion. The dynamics on $\cH_{d-1}$ is non-chaotic 
and the spacetime metric asymptotically tends 
to a generalized Kasner metric, corresponding 
to an uninterrupted geodesic motion of the ball.

The question of chaos vs. regular motion is thereby reduced
to determining whether it is case 1 or case 2 that is realized. 
We discuss this in the next section by relating the
``wall cone" ${\cal W}^+$ to the fundamental Weyl chamber of
a certain indefinite Kac-Moody algebra.

\section{Hyperbolic Kac-Moody algebras and chaos}

In this section, we show that the
reflections (\ref{coll}) can be identified with 
the fundamental Weyl reflections of the indefinite Kac-Moody
algebra $AE_d$, and therefore that
the cone ${\cal W}^+$ can be identified with
the fundamental Weyl chamber of $AE_d$. To do that, we 
need to compute the dihedral angles between the walls.
A direct calculation shows that the Gram matrix 
\begin{equation}\label{Gram}
A_{ij}
\equiv G^{\mu\nu}w_{i\mu} w_{j\nu} \qquad
\hbox{for $i,j = -1, 0, 1, \cdots, n$} 
\end{equation}
of the scalar products of the wall forms is given by
\begin{equation}
\hspace{1.5cm} A_{ij} = \left( \matrix{2 &-1 & 0 \cr 
-1 & 2 & -2 \cr 
0 & -2 & 2  \cr
} \right) \, \; \; \; \; \hbox{for $d=3$}
\label{gram1}
\end{equation} 
and
\begin{equation}
A_{ij} = \left( \matrix{2 &-1 & 0 & 0 & \cdots & 0 & 0 & 0 \cr
-1 & 2 &-1 & 0 &  \cdots & 0 & 0 & -1\cr
0 & -1 & 2 &-1 & \cdots & 0 & 0 & 0 \cr
\vdots & & & & & & & \cr
0 &0 & 0 & 0 & \cdots &2& -1 & 0\cr
0 &0 & 0 & 0 & \cdots &-1 & 2 &-1 \cr
0 & -1 & 0 & 0 & \cdots & 0 & -1 & 2 \cr
} \right) \, \; \; \; \; \hbox{for $d >3$}.
\label{gram2}
\end{equation}
In both cases, the wall forms have same length $\sqrt{2}$. As in 
\cite{tDmH}, we identify them with the simple roots of a Kac-Moody
algebra. To emphasize the identifications ``wall forms = simple
roots", we shall henceforth switch to a new notation and denote the wall
forms $w_i$ by $r_i$.  We shall also denote the Cartan subalgebra of
the Kac-Moody algebra by {\goth H} and its dual 
(space of linear forms on {\goth H}, i.e., the
``root space") by {\goth H}$^*$.
Thus, 
\begin{equation}
w_i \equiv r_i \in \hbox{{\goth H}}^*.
\label{iden1}
\end{equation}
We recall that the root space {\goth H}$^*$ 
is endowed with a bilinear form,
which we identify with the bilinear form defined
by the (contravariant) metric $G^{\mu \nu}$ given above,
\begin{equation}
r_i \cdot r_j \equiv G^{\mu \nu} r_{i \mu} r_{j \nu}
\label{iden2}
\end{equation} 
Since the roots have all same length 
squared $2$, the algebra is ``simply-laced" and the Gram matrix $A_{ij}$
computed in (\ref{gram1}) and (\ref{gram2}) is 
also the Cartan matrix $a_{ij}$,
\begin{equation}
a_{ij} \equiv \frac{2 r_i \cdot r_j}{r_i \cdot r_i}, 
\end{equation}
i.e., $A_{ij} = a_{ij}$. 
We then recognize the first matrix as the 
Cartan matrix of the Kac-Moody algebra $AE_3$, while
the second matrix is the Cartan matrix of the Kac-Moody algebra
$AE_d$ ($d>3$). This is what justifies the identifications
(\ref{iden1}) and (\ref{iden2}).
The roots $r_0, \dots r_n$ form the closed 
ring of the Dynkin diagram, $r_0$ is the (affine) root closing the 
ring, and $r_{-1}$ is the overextended root connected to $\R_0$. 

Once the wall forms are identified with the simple roots of a Kac-Moody
algebra, the space $\cM_d$ in which the dynamics of the scale
factors takes place becomes identified with the Cartan subalgebra
{\goth H} of $AE_d$.  The cone
${\cal W}^+$ defining the billiard is given by the conditions
\be
\langle \R_i, \beta \rangle  \; \;  \geq 0
\qquad \hbox{for all $i=-1,0,1,...,n$}
\label{W+}
\ee
where $\langle \R_i, \beta  \rangle$ denotes the pairing between a form $r_i 
\in$ {\goth H}$^*$ and a vector $\beta \in$ {\goth H}.
The cone ${\cal W}^+$ is 
then just the fundamental Weyl chamber \cite{Kac,MP}, as was 
anticipated by our notations. It is striking to note that 
the finite dimensional germ $A_{d-2}$ of the hyperbolic algebra 
$AE_d$ is nothing but the Ehlers symmetry of the toroidal 
compactification of the original gravity  from $d+1$ to 3 dimensions 
\cite{CJLP3}. The reduction  to two dimensions brings the affine extension 
and the final elimination of all spatial coordinates increases the rank 
further to $d$ \cite{julia80}. 
 
The above Cartan matrices are indecomposable.  They are also
of indefinite, Lorentzian
type since the metric $G_{\mu \nu}$ in {\goth H} is of Lorentzian signature.  
A Cartan matrix with these properties is said to be of hyperbolic 
type if any subdiagram obtained by removing a node from its
Dynkin diagram is either of finite or affine type \cite{Kac}.
The concept of hyperbolicity is particularly relevant here because 
it is a general result that the fundamental Weyl chamber ${\cal W}^+$ 
of a hyperbolic Kac-Moody algebra is contained within the 
light cone ${\cal C}^+$; the Weyl cell is then a (generalized) 
simplex of finite volume. Furthermore, for hyperbolic KM algebras
the closure of the Tits cone, defined as the union of the 
fundamental Weyl chamber and all its images under the Weyl
group, is just ${\cal C}^+$ (\cite{Kac}, section 5.10).  

As already mentioned, the Kac-Moody algebras $AE_d$
are hyperbolic for $d \leq 9$. 
We will now verify by explicit computation 
that their associated fundamental Weyl chambers are indeed contained 
in the forward light cone. The location of the fundamental Weyl 
chambers in the general case is most conveniently (and most easily)  
analyzed by means of the fundamental weights
$\GL_j \in$ {\goth H}$^*$. The latter are defined by
\be
\R_i \cdot \GL_j \,  \equiv \, 
G^{\mu\nu}  \R_{i\mu}  \GL_{j\nu} = \delta_{ij} \; \; \; \; \; 
i,j=-1,0,1,\dots,n \equiv d-2.
\label{weight}
\ee    
Let us also introduce the coweights $\Lambda_i^\vee \in$ {\goth H},
i.e., the contravariant vectors associated with the forms
$\Lambda_i$ with components
$(\Lambda_i^\vee)^\mu \equiv G^{\mu \nu} \Lambda_{i \nu}$.
Because the fundamental Weyl chamber ${\cal W}^+$ is defined by the
conditions $\langle \R_i,  \beta \rangle \geq 0$, we have
\be
{\cal W}^+ = \Big\{ \beta \in \cM_d \equiv \hbox{{\goth H}} \; \big|\;
\beta = \sum_{i=-1}^n a_i \Lambda_i^\vee \;\; , \;\; a_i \in \BR, \;
a_i \geq 0 \Big\}
\ee
The (one-dimensional) 
edges of ${\cal W}^+$ are obtained by setting all $a_j$ 
except one to zero, which gives the vectors $\Lambda_i^\vee$.  
The question of determining whether the fundamental Weyl 
chamber is contained in the forward light cone or not is 
thus reduced to a simple computation of the norms of 
the fundamental weights. 

To get the fundamental weights, we observe that if the root $\Rm$ is 
dropped, the associated Cartan matrix reduces to the Cartan matrix of 
affine $sl(n+1)$. The affine null root is given by
\be
\delta = \R_0 + \R_1 + ... \R_n
\ee
It obeys $\Gd^2 \equiv \Gd \cdot \Gd = 0 
= \R_j \cdot \Gd $ for all 
$j=0,1,...,n$ (but $\Rm \cdot \Gd  = -1$).
The fundamental weights for the subalgebra $A_n$ are defined by
\be
\R_i \cdot \Gl_j = \delta_{ij} \qquad 
      {\rm for} \;\; i,j=1,...,n
\ee
They are explicitly given by
\ba
\Gl_j &=& 
\frac{n-j+1}{n+1} \Big[ \R_1 + 2 \R_2 + \dots + j \R_j \Big] \\
&& + \frac{j}{n+1} \Big[ (n-j) \R_{j+1} + (n-j-1) \R_{j+2} + \dots
   + \R_n \Big]
\ea
with norm
\be
\Gl_j^2 = \frac{j(n-j+1)}{n+1} > 0
\ee
(note that $\R_0 \cdot \Gl_j  = -1$ for all $j=1,...,n$).
One then finds for the fundamental weights\footnote{In the 
general case with highest root ${\bf \theta}= \sum_j m_j \R_j$, 
we have $\R_0 \cdot \Gl_j  = -m_j$ and the
fundamental weights are given by
$$\GL_{-1} = - \Gd \; , \quad \GL_0 = -\Rm - 2\Gd \; , \quad
\GL_j = m_j \GL_0 + \Gl_j
$$
An alternative representation is $\GL_i = \sum_j (a^{-1})_{ij} \R_j$
where $(a^{-1})_{ij}$ is the inverse Cartan matrix.} of $AE_d$
\begin{equation}
\GL_{-1} = - \Gd \; ,   \quad 
\GL_0 = - \Rm - 2\Gd \; ,  \quad 
\GL_j =  \GL_0 + \Gl_j \; \; {\rm for} \;\; j=1,...,n
\end{equation}
Their norms (with $\GL^2 \equiv \GL \cdot \GL \equiv
G^{\mu\nu} \GL_\mu \GL_\nu$) are 
easily computed:
\begin{equation}
\GL_{-1}^2 = 0\; ,  \quad
\GL_0^2 = -2\; ,  \quad
\GL_j^2 = -2 + \frac{j(n-j+1)}{n+1} 
\end{equation}
Note that $\GL_{-1}$ is always lightlike, and $\GL_0$ is timelike 
for all $n$. It is furthermore elementary to check that
\be
\GL_j^2 \leq 0 \qquad {\rm for}\; {\rm all} \; j \;\; 
            {\rm if} \;\; n\leq 7
\ee
with equality only for $n=7$ and $j=4$. For $n\geq 8$ there is 
always at least one {\em spacelike} fundamental weight $\GL_j$; 
e.g. for $n=8$ we have
\be
\GL_4^2 = \GL_5^2 = \frac29 > 0
\ee 

The above calculation then tells us that for $n\leq 7 $ 
(i.e. for $AE_d$ with $d\leq 9$) the fundamental
Weyl chamber is contained in the forward light cone with 
one edge touching the light cone (two edges for $n=7$). 
For $n\geq 8$ there is at least one spacelike edge, so the Weyl 
chamber contains timelike, lightlike and spacelike vectors.
This is, then, the Kac-Moody theoretic understanding of the fact
that the asymptotic solution of the vacuum Einstein equations 
in the vicinity of a spacelike singularity exhibits the 
never-ending oscillatory behaviour of the BKL type in 
spacetime dimensions $\leq 10$, while this ceases to be the
case for $D \geq 11$ \cite{DHS}.  

To conclude this letter we would like to stress once more
that the emergence of a Kac-Moody algebra is not automatic for
the gravitational systems under consideration. For instance,
the billiard associated with the Einstein-Maxwell system in $D$ 
spacetime dimensions has the same symmetry walls (\ref{symmetry1}),
(\ref{symmetry2}), (\ref{symmetry3}), but the gravitational
wall (\ref{gravitational}) is replaced by the (asymptotically
dominant) 
electric wall $w_1(\beta) = \beta^1$. This wall
is orthogonal to
all symmetry walls, except $w_2$ ($w_0$ for $d = 3$) with which 
it makes an angle $\alpha$ given by 
$\cos \alpha =  \sqrt{(d-1)/2(d-2)}$. This dihedral angle is 
generically not a submultiple of $\pi$ and the associated group 
of reflections is not a discrete group, with two notable exceptions:
($i$) $\alpha$ is equal to zero for $D=4$, where electric and 
gravitational walls coincide (though the wall forms are normalized 
differently), and ($ii$) the angle $\alpha$ is equal to $\pi/6$
for the case $D=5$, whose study was advocated in \cite{dh2}
in the context of homogeneous models. 
Taking into account that the wall form $w_1$ has norm squared equal to
$(d-2)/(d-1) = 2/3$, one gets in that case the Cartan matrix
\begin{equation}
\hspace{1.5cm} a_{ij} = \left( \matrix{2 &-1 & 0 &0 \cr
-1 & 2 & 0 & -1\cr
0 & 0 & 2  &-3\cr
0 & -1 & -1 &2 \cr
} \right) 
\label{cartan}
\end{equation}      
The underlying Kac-Moody
algebra is the canonical hyperbolic extension of
the exceptional Lie agebra $G_2$ (hyperbolic algebra number
16 in table 2 of \cite{Saclio}).  One Einstein-Maxwell theory
in $5$ dimensions is particularly interesting because it
is the bosonic sector of simple supergravity in $5$ dimensions,
which shares many similarities with $D=11$ supergravity, such as
the cubic Chern-Simons term for the vector field \cite{ChNi}.
The relevance of the exceptional group $G_2$ to that theory was 
pointed out in \cite{Mizo,CJLP3}. This system,
as well as pure gravity or superstring models 
and $M$-theory, for which one does get the Weyl group of a 
Kac-Moody algebra, are thus rather 
exceptional \cite{tDmH}.

\section*{Acknowledgements}
T.D. thanks Victor Kac for informative communications.
M.H. and H.N. are grateful to the Institut des Hautes Etudes
Scientifiques for its kind hospitality.
The work of M.H. is partially supported by the ``Actions de
Recherche Concert{\'e}es" of the ``Direction de la Recherche
Scientifique - Communaut{\'e} Fran{\c c}aise de Belgique", by
IISN - Belgium (convention 4.4505.86) and by the European Commission 
programme HPRN-CT-2000-00131 in which he is associated to K.U.Leuven. 
The work of H.N. was supported in part by the European Union 
under Contract No. HPRN-CT-2000-00122.  The work of B.J. 
was supported in part by the European Union
under Contract No. HPRN-CT-2000-00131.

\end{document}